\documentclass[12pt]{article}
\usepackage{amsmath}
\usepackage{amsthm}
\usepackage{amssymb}
\usepackage{graphicx,psfrag,epsf}
\usepackage{enumerate}
\usepackage{natbib}
\usepackage{subfigure}
\usepackage{multirow}

\newcommand{\blind}{0}

\newcommand{\M}[1]{\boldsymbol{#1}}  
\newcommand{\V}[1]{\boldsymbol{#1}}  
\newcommand{\Unif}[0]{\textrm{Uniform}}

\DeclareMathOperator*{\argmin}{\textrm{argmin}}

\begin{document}

\def\spacingset#1{\renewcommand{\baselinestretch}%
{#1}\small\normalsize} \spacingset{1}


\if0\blind
{
  \title{\bf Directional Statistics of Preferential Orientations of Two Shapes in Their Aggregate and Its Application to Nanoparticle Aggregation}
  \author{Ali Esmaieeli Sikaroudi, David A. Welch, Taylor J. Woehl, Roland Faller, \\
  James E. Evans, Nigel D. Browning \& Chiwoo Park \thanks{
      Ali Esmaieeli Sikaroudi (ali.esmaieeli@gmail.com) is Business Analyst in JPMorgan Chase, Jacksonville, FL 32256. David A. Welch (dawelch@ucdavis.edu) is Adjunct Instructor at Suffolk County Community College, and Roland Faller (rfaller@ucdavis.edu) is Professor in the Department of Chemical Engineering, University of California at Davis, One Shields Avenue, Davis, CA 95616. Taylor Woehl (tjwoehl@umd.edu) is Assistant Professor in the Department of Chemical and Biomolecular Engineering, University of Maryland, College Park, MD 20742. James Evans (james.evans@pnnl.gov) is Scientist III at Environmental Molecular Sciences Laboratory, Pacific Northwest National Laboratory, 902 Battelle Blvd, Richland, WA 99354. Nigel Browning (nigel.browning@pnnl.gov) is Chief Scientist at Fundamental Computational Sciences Directorate, Pacific Northwest National Laboratory, 902 Battelle Blvd, Richland, WA 99354. Chiwoo Park (cpark5@fsu.edu) is Associate Professor in the Department of Industrial and Manufacturing Engineering, Florida State University, Tallahassee, FL 32310. We acknowledge support through the Laboratory Directed Research and Development (LDRD) Program Chemical Imaging Initiative at Pacific Northwest National Laboratory (PNNL) as performed in the Environmental Molecular Sciences Laboratory (EMSL), a national scientific user facility sponsored by DOE's Office of Biological and Environmental Research at PNNL. PNNL is a multiprogram national laboratory operated by Battelle for the United States Department of Energy (DOE) under Contract DE-AC05-76RL01830. This work is also supported by the National Science Foundation under NSF-1334012, the Air Force Office of Scientific Research under FA9550-13-1-0075, and FSU PG 036656. The last author is the corresponding author.}
    }
  \maketitle
} \fi

\if1\blind
{
  \bigskip
  \bigskip
  \bigskip
    \title{\bf Directional Statistics of Preferential Orientations of Two Shapes in Their Aggregate and Its Application to Nanoparticle Aggregation}
  \maketitle
} \fi

\bigskip
\begin{abstract}
Nanoscientists have long conjectured that adjacent nanoparticles aggregate with one another in certain preferential directions during a chemical synthesis of nanoparticles, which is referred to the oriented attachment. For the study of the oriented attachment, the microscopy and nanoscience communities have used dynamic electron microscopy for direct observations of nanoparticle aggregation and have been so far relying on  manual and qualitative analysis of the observations. We propose a statistical approach for studying the oriented attachment quantitatively with multiple aggregation examples in imagery observations. We abstract an aggregation by an event of two primary geometric objects merging into a secondary geometric object. We use a point set representation to describe the geometric features of the primary objects and the secondary object, and formulated the alignment of two point sets to one point set to estimate the orientation angles of the primary objects in the secondary object. The estimated angles are used as data to estimate the probability distribution of the orientation angles and test important hypotheses statistically. The proposed approach was applied for our motivating example, which demonstrated that nanoparticles of certain geometries have indeed preferential orientations in their aggregates.
\end{abstract}

\noindent%
\vfill

\newpage
\spacingset{1.45} 

\section{INTRODUCTION} \label{sec:intro}
A particle aggregation is a merging of two smaller particles into one larger particle, which is one of the main driving forces that grow atoms or molecular clusters into nanoparticles during a chemical synthesis of nanoparticles. With a better understanding of a particle aggregation, synthesizing nanoparticles of desired sizes and shapes should be possible \citep{welch2015nature, zhang2012attachment, li2012direction}. 

\begin{figure}[b]
	\centering
		\includegraphics[width=0.8\textwidth]{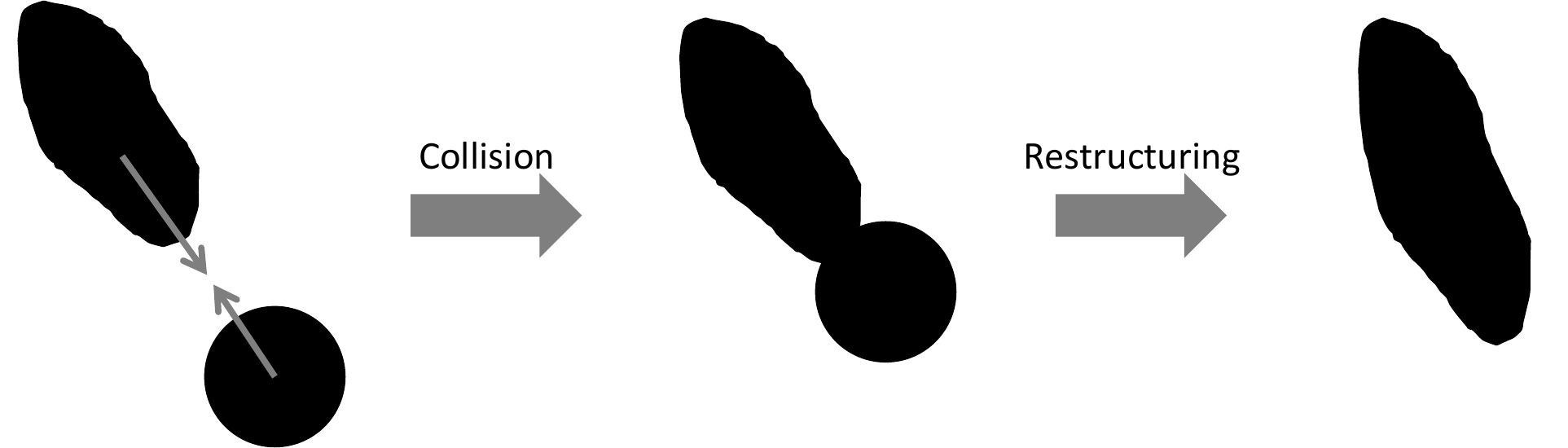}
	\caption{A picture illustrating a particle aggregation. A particle aggregation is a collision of two primary particles followed by a restructuring of the collision outcome to a secondary particle}
	\label{fig:fig1}
\end{figure}

As seen in Figure \ref{fig:fig1}, a particle aggregation is essentially a two-step process, a collision of two primary particles followed by their restructuring to a larger secondary particle. Some collisions effectively lead to a subsequent restructuring  (or coalescence), while other collisions are ineffective. The degree of effectiveness depends on how primary nanoparticles are spatially oriented in a collision. When primary particles are oriented ineffectively, they become separate again or rotate to a preferred orientation, as in the phenomenon known as \textit{the oriented attachment} \citep{li2012direction}. An open research problem concerns the oriented attachment. The physical phenomenon can be directly observed using a state-of-the-art electron microscope, but analyzing a number of nanoparticle aggregation cases appearing in microscope images remains challenging. This paper addresses how to study the microscopic observations of nanoparticle aggregations to statistically analyze the preferential orientations of primary nanoparticles. 

A major contribution of this paper is to provide mathematical and statistical modelings for accurately studying the oriented attachment of primary nanoparticles.  The microscopy and nanoscience communities have been relying on manual analysis of a very few examples of nanoparticle aggregation for the study of the oriented attachment. Our proposed method will provide a systematic way of statistically analyzing a large population of aggregation examples to find a statistically reliable estimate of the preferential orientations of nanoparticles within their aggregations. We acknowledge that there are some existing works on the statistical analysis of aggregates in concrete and asphalt engineering \citep{mora2000sphericity, wang1999image}, but those works primarily focused on studying how aggregation outcomes are sized and shaped, instead of studying how aggregating components are oriented. We believe that our work is the first of its kind in statistically studying the oriented attachment of nanoparticles. 

In addition to the contribution in applications and modeling, this paper contains two methodological contributions. In statistical shape analysis, a problem of aligning one shape to another shape has been well studied to possibly find the relative orientation of one to another \citep{schmidler2007fast, green2006bayesian}. However, the existing theory and methods do not work for analyzing the orientations of two aggregating components within their aggregate, which involves aligning two shapes to one shape that is assumed to be a union of the two shapes. This paper presents in Section \ref{sec:align} an approach to this two-to-one alignment problem for two-dimensional shapes. The accuracy of the proposed approach was evaluated in Section \ref{sec:simu} using an example system of ellipse aggregations. The approach may be applied for analyzing aggregations of other shapes with additional numerical validations. On the other hand, in directional statistics, angular data and their distribution have long been studied \citep{fisher1995statistical}, but studies on the probability distribution of angular data with some symmetries are lacking. For our motivating example, the angular distribution of a particle orientation is essentially four-fold symmetric due to geometrical symmetries of nanoparticles. Section \ref{sec:stat} presents a new probability distribution to model the four-fold symmetry and the related statistical analysis.

The remainder of this paper is organized as follows. Section \ref{sec:dataset} describes microscopy data that motivated this study. Section \ref{sec:model} presents how we mathematically model an aggregate and the orientations of aggregating components. Section \ref{sec:stat} describes several statistical inference problems on the orientations, including a probability density estimation problem and some statistical hypothesis testing problems, which were applied in Section \ref{sec:exp} to test several scientific hypotheses posed to explain the oriented attachment. Section \ref{sec:concl} provides our conclusions.

\section{DATASET} \label{sec:dataset}
\begin{figure}[t]
	\centering
		\includegraphics[width=\textwidth]{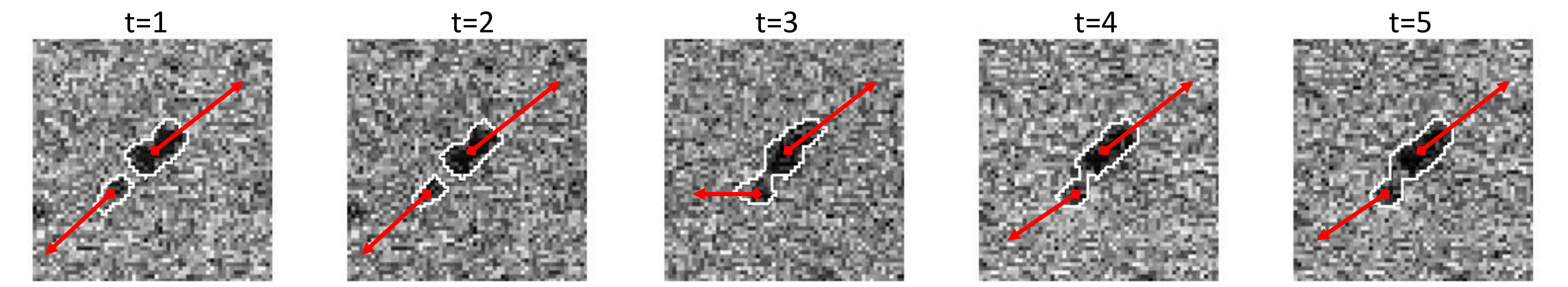}
	\caption{Example dynamic microscopy data of particle aggregation. This example consists of a sequence of five microscope images that show the movement and aggregation of two nanoparticles. The second image labeled `t=2' is the image of the two nanoparticles taken immediately before the aggregation, and the fourth image labeled `t=4' is the image of the aggregation outcome.}
	\label{fig:fig3}
\end{figure}

We used dynamic scanning transmission electron microscopy to synthesize and directly observe growth of silver nanoparticles \citep{taylor12}, taking a sequence of electron microscope images of about two hundred silver nanoparticles and their aggregations. We applied an object-tracking algorithm \citep{parkpami2013} with the microscope images to track their aggregations, which identified 184 different aggregation cases. Figure \ref{fig:fig3} displays an example of the captured aggregation events. 

For each aggregation event, we take two items of information: the first is the image of two primary nanoparticles taken immediately before the aggregation, e.g., the image at $t=2$ in Figure \ref{fig:fig3}, and the second is the image of the secondary nanoparticle taken immediately after the final aggregation, e.g. the image at $t=4$. After the final aggregation, the orientations of the two primary nanoparticles do not change due to strong physical forces as shown in Figure \ref{fig:fig3}. Therefore, the aggregate image can be taken any time after the final aggregation, but our choice is the time immediate after the aggregation because the aggregate might later undergo a significant restructuring. The time resolution of the imaging process is faster than a normal aggregation speed, so the `immediate before the aggregation' and the `immediate after the aggregation' are well defined from the observed image sequences. 

Each of the before images and the after images is two-dimensional, depicting the projection of the three dimensional geometries of nanoparticles on a two-dimensional space. Since the nanoparticles imaged are constrained to a very thin layer of a sample chamber, we assume that geometrical information along the $z$-direction is relatively insignificant. A set of the image pairs for the 184 aggregation events will be analyzed for studying how primary nanoparticles are oriented in their aggregates. 

\section{MODELING AGGREGATION} \label{sec:model}
We abstract an aggregation as a merge of two geometric objects. We first describe how we model geometric objects. Let $\mathbb{X}$ denote a set of all image pixel coordinates in an $H \times W$ two-dimensional digital image, 
\begin{equation*}
\mathbb{X} := \{ (h, w): h = 0, 1, 2,...,H, w = 0, 1, 2,...,W \}.
\end{equation*}
A geometric object imaged on $\mathbb{X}$ is represented by a simply connected subset of $\mathbb{X}$ that represents a set of all image pixel coordinates locating inside the geometric object. The set-based representation has been popularly used for shape analysis \citep{memoli2005theoretical, memoli2007use}, which seems more useful for our motivating problem than other popular shape representation models such as the representation by landmark points \citep{kendall1984shape, dryden1998statistical} and the representation by a closed curve \citep{younes1998computable, srivastava2011shape}. The landmark-based approach has a major technical issue regarding how to manually choose the landmarks of many geometrical bodies, which are also subject to human bias. More importantly, an aggregation of two geometric objects is better represented by the set-based representation. An aggregation of two objects can be naturally represented by the union of two subsets representing the two objects. 

\begin{figure}[t]
	\centering
		\includegraphics[width=0.8\textwidth]{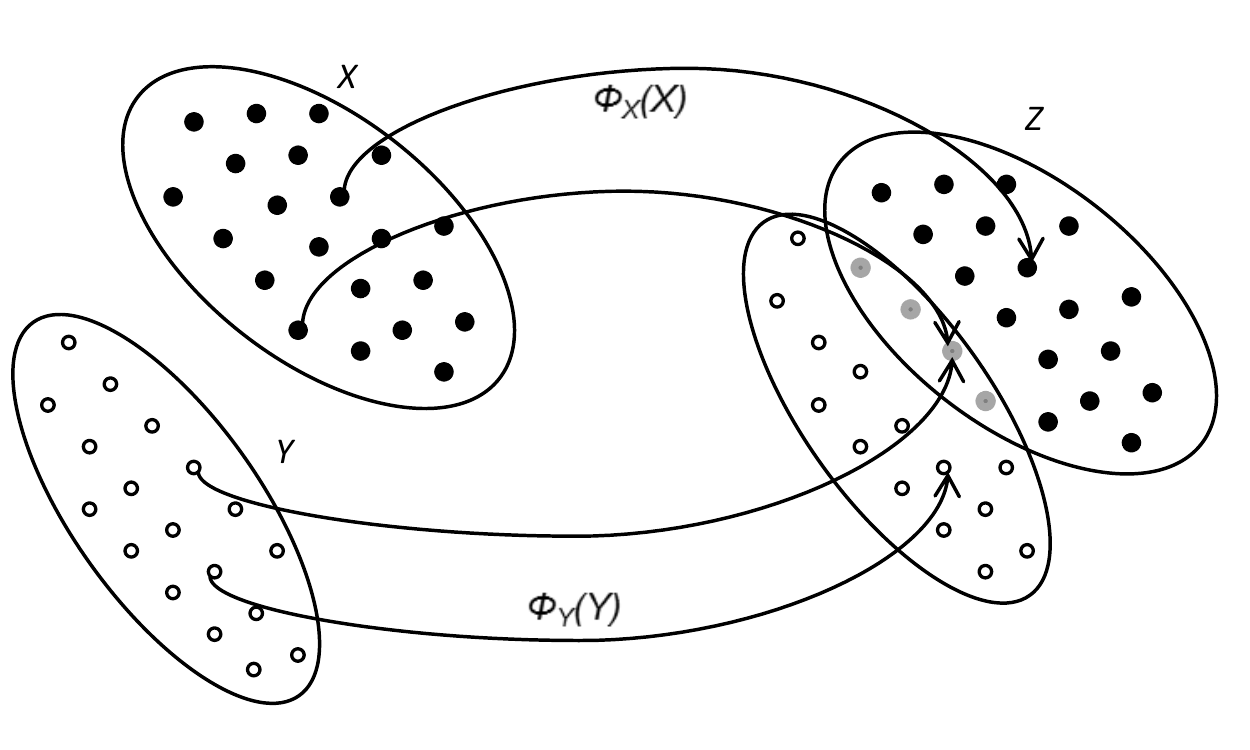}
	\caption{Set-based representation of two particles and a particle aggregate. $X$ and $Y$ are two simply connected sets that represent the two particles, and $Z$ is an aggregation of the two particle, represented by an union of $X$ and $Y$ transformed by the Euclidean rigid body transformations $\phi_X$ and $\phi_Y$.}
	\label{fig:fig2}
\end{figure}

Geometric objects move and rotate before they aggregate. The movement and rotation operations in $\mathbb{X}$ are represented by a Euclidean rigid body transformation. Let $\mathbb{SE}(\mathbb{X})$ denote a collection of all Euclidean rigid body transformations defined on $\mathbb{X}$. An element $\phi$ in $\mathbb{SE}(\mathbb{X})$ is an Euclidean rigid body transformation that basically shifts $\V{x} \in \mathbb{X}$ by $\V{c}_{\phi} \in \mathbb{X}$ in the negative direction and rotates the shifting result about the origin by $\theta_{\phi} \in [0, 2\pi]$,  
\begin{equation} \label{eq:phi}
\phi(\V{x}) = \left[ \begin{array}{c c} \cos(\theta_{\phi}) & -\sin(\theta_{\phi}) \\ \sin(\theta_{\phi}) & \cos(\theta_{\phi}) \end{array} \right] (\V{x} - \V{c}_{\phi}).
\end{equation}
The $\V{c}_{\phi}$ is referred to as the translation vector of $\phi$, and the $\theta_{\phi}$ is referred to as the rotation angle of $\phi$. 
For a set $X \subset \mathbb{X}$ and a transformation $\phi \in \mathbb{SE}(X)$, we use a notation $\phi(X)$ to denote the image of $X$ transformed by $\phi$, 
\begin{equation*}
\phi(X) = \{\phi(\V{x}); \V{x} \in X\}.
\end{equation*}
When $X$ represents a geometric object, $\phi(X)$ represents the image of the geometric object transformed by the movement and rotation operations defined by $\phi$. The operations do not deform the geometric object but just change its configuration parameters, i.e., locations and orientations, which is why $\phi$ is called a rigid body transformation.  

Let $X \subset \mathbb{X}$ and $Y \subset \mathbb{X}$ denote two simply connected subsets of $\mathbb{X}$ that represent two primary objects, and let $Z \subset \mathbb{X}$ denote a simply connected subset of $\mathbb{X}$ that represents the aggregate of the two primary objects. The two primary objects may move and rotate before they collide and aggregate. Let $\phi_X \in \mathbb{SE}(\mathbb{X})$ and $\phi_Y \in \mathbb{SE}(\mathbb{X})$ denote the Euclidean rigid body transformations that represent the movements and rotations of $X$ and $Y$ before they aggregate. Let $\V{c}_{\phi_X}$ and $\V{c}_{\phi_Y}$ denote the translation vectors of the two transformations, and let $\theta_{\phi_X}$ and $\theta_{\phi_Y}$ denote the rotation angles of the transformations. As shown in Figure \ref{fig:fig2}, before the aggregate $Z$ is fully restructured to a different shape, $Z$ is approximately an overlapping union of $\phi_X(X)$ and $\phi_Y(Y)$,
\begin{equation*}
Z = \phi_X(X) \cup \phi_Y(Y).
\end{equation*} 
In practice, $\mathbb{X}$ is a digital image, so the equality does not exactly hold due to digitization errors. The aggregate $Z$ can be partitioned into three pieces, $Z_1 = \phi_X(X) \backslash \phi_Y(Y)$, $Z_2 =\phi_Y(Y) \backslash  \phi_X(X)$ and $Z_3 = \phi_X(X) \cap \phi_Y(Y)$, where $\backslash$ is a set difference operator. We call the center of mass of $Z_3$ as an aggregation center, which we denote by $\V{c}_{X,Y}$. 

\begin{figure}[h]
	\centering
		\includegraphics[width=0.6\textwidth]{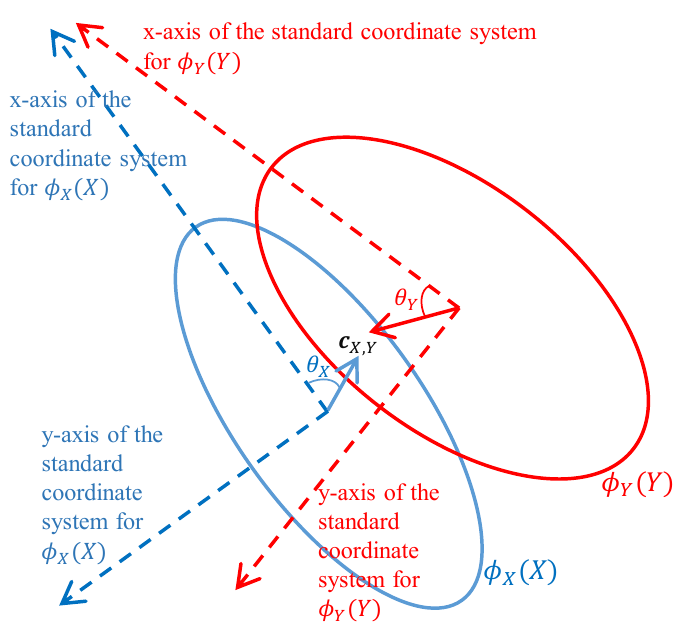}
	\caption{Definition of $\theta_X$ and $\theta_Y$. The symbol $\V{c}_{X, Y}$ represents the center of the intersection of two aggregating components, so it belongs to a part of $\phi_X(X)$ and also belongs to a part of $\phi_Y(Y)$. The $\theta_X$ basically defines which part of $\phi_X(X)$ intersects with $\phi_Y(Y)$, while the $\theta_Y$ defines which part of $\phi_Y(Y)$ intersects with $\phi_X(X)$.}
	\label{fig:fig21}
\end{figure}

We define the orientation of $X$ in $Z$ as the orientation of $\V{c}_{X,Y}$ in the standard coordinate system of $\phi_X(X)$ as described in Figure \ref{fig:fig21}. Let us represent the standard coordinate system for $X$ by an one-to-one map $T_X: \mathbb{X} \rightarrow \mathbb{R}^2$ that assigns to a point $x \in X$ a pair of numerical coordinates. The map $T_X \circ \phi_X^{-1}$ defines the standard coordinate system for $\phi_X(X)$ induced by $T_X$, because for $y \in \phi_X(X)$, $\phi_X^{-1}(y) \in X$ is uniquely determined since $\phi_X$ is a bijection, and $T_X$ can assign to the point $\phi_X^{-1}(y) \in X$ a pair of the unique coordinate numbers, $T_X \circ \phi_X^{-1}(y)$. Therefore, the orientation of $X$ in $Z$ is 
\begin{equation*}
\V{v}_X = \frac{T_X \circ \phi_X^{-1} (\V{c}_{X,Y})}{||T_X \circ \phi_X^{-1} (\V{c}_{X, Y})||} \mbox{, or } \theta_X = \mbox{angle}(\V{v}_X),
\end{equation*} 
where angle$(\V{v}_X)$ is the angular part of the polar coordinate of $\V{v}_X$. Similarly, the orientation of $Y$ in $Z$ is defined by 
\begin{equation*}
\V{v}_Y = \frac{T_Y \circ \phi_Y^{-1} (\V{c}_{X,Y})}{||T_Y \circ \phi_Y^{-1} (\V{c}_{X, Y})||} \mbox{, or } \theta_Y = \mbox{angle}(\V{v}_Y).
\end{equation*}
Our primary interest is to study the oriented attachment, i.e., investigating what angles of $\theta_X$ and $\theta_Y$ are more frequently observed from multiple aggregation examples. 

The $T_X$ and $T_Y$ are independent of $\phi_X$, $\phi_Y$ and $\V{c}_{X,Y}$, i.e., the choice of the former does not affect the latter, and vice versa. Therefore, estimating $T_X$ and $T_Y$ can be performed independently to estimating the others. Section \ref{sec:align} defines the parametric forms of $T_X$ and $T_Y$ and describes how to estimate their parameters. Section \ref{sec:aggregate} describes how to estimate $\V{c}_{X,Y}$, the parameters of $\phi_X$ (i.e. $\V{c}_{\phi_X}$ and $\theta_{\phi_X}$), and the parameters of $\phi_Y$ (i.e. $\V{c}_{\phi_Y}$ and $\theta_{\phi_Y}$). The accuracy of the estimation are validated using simulation datasets in Section \ref{sec:simu}.

\subsection{Estimation of $T_X$} \label{sec:align}
The standard coordinate system of $X$ must be consistently defined with those of other geometric objects geometrically similar to $X$, so their orientations can be defined consistently. To accomplish this, we define a reference shape for a collection of geometric objects geometrically similar to $X$ and define $T_X$ as the Euclidean rigid body transformation that best aligns $X$ to the reference shape. The transformation outcome is invariant to a Euclidean rigid body transformation of $X$, i.e., $T_X(X) = T_{\phi(X)}(\phi(X))$ for $\phi \in \mathbb{SE}(\mathbb{X})$, unless the reference shape is redefined, so it provides consistent coordinate numbers for those having similar geometries but different orientations and locations. In this section, we describe how we define a reference shape, and our proposal for estimating $T_X \in \mathbb{SE}(\mathbb{X})$ given a reference shape is described.

We first work on how to estimate $T_X$ when a reference shape is given. Let $X$ and $X_0$ denote the simply connected subsets of $\mathbb{X}$ that represent a geometric object and its reference shape respectively. Suppose that $X$ and $X_0$ consist of $m$ and $m_0$ point coordinates as follows,
\begin{equation*}
\begin{split}
& X = \{\V{x}_1, \V{x}_2, \ldots, \V{x}_{m}\} \mbox{ and } \\
& X_0 = \{\V{x}^{(0)}_1, \V{x}^{(0)}_2, \ldots, \V{x}^{(0)}_{m_0}\},
\end{split}
\end{equation*} 
where $\V{x}_i \in \mathbb{X}$ denotes the $i$th element of $X$, and $\V{x}^{(0)}_j \in \mathbb{X}$ indicates the $j$th element of $X_0$. We want to find $T_{X} \in  \mathbb{SE}(\mathbb{X})$ that best aligns $X$ to $X_0$,  
\begin{equation*}
T_{X}(X) \approx X_0,
\end{equation*}
where the closeness of the two sets is measured by a set distance. A popular set distance is the $p$ norm distance \citep{memoli2007use}, which basically averages the distances between each pair of the elements in the two sets that correspond to each other. Let $\mu_{ij}$ define the following measure of correspondence between the elements of the two sets, $T_{X}(X)$ and $X_0$,
\begin{equation} \label{eq:mu}
\mu_{ij} = 1 \mbox{ if } T_{X}(\V{x}_i) \mbox{ corresponds to } \V{x}^{(0)}_j \mbox{ and } 0 \mbox{ otherwise}.
\end{equation}
When $\mu_{ij}$'s are known, the set distance is defined by 
\begin{equation*}
dist(T_X(X), X_0; \V{\mu}) = \left(\sum_{i,j} \mu_{ij} \left|\left|T_X(\V{x}_i) - \V{x}^{(0)}_j\right|\right|^p\right)^{1/p},
\end{equation*}
where $\M{\mu}$ denotes a $m \times m_0$ matrix with the $(i,j)$th element $\mu_{ij}$. The $T_X$ that best aligns $X$ to $X_0$ can be achieved by minimizing the distance,
\begin{equation*}
T_{X}^*(\V{\mu}) = \argmin_{T_X \in \mathbb{SE}(\mathbb{X})} dist(T_X(X), X_0; \V{\mu}).
\end{equation*}
Let $\V{c}_{T_X}^*$ and $\theta^*_{T_X}$ denote the translation vector and rotation angle of $T_{X}^*(\V{\mu})$. The expression for the two parameters can be found at \citet{rangarajan1997softassign}, 
\begin{equation} \label{eq:rigid}
\begin{split}
& \V{c}_{T_X}^* = \frac{\sum_{i=1}^{m} \sum_{j=1}^{m_0} \mu_{ij} (\V{x}_i - \V{x}^{(0)}_j)}{\sum_{i=1}^{m} \sum_{j=1}^{m_0} \mu_{ij}} \mbox{ and }\\
& \theta^*_{T_X} = \arctan\left(\frac{\sum_{i=1}^{m} \sum_{j=1}^{m_0} \mu_{ij} (\V{x}^{(0)}_j \times \V{x}_i)}{\sum_{i=1}^{m} \sum_{j=1}^{m_0}  \mu_{ij} (\V{x}^{(0)}_j \cdot \V{x}_i) }\right),
\end{split}
\end{equation} 
where $(a_1, a_2) \times (b_1, b_2) = a_1b_2 - a_2 b_1$ and $(a_1, a_2) \cdot (b_1, b_2) = a_1b_1 + a_2 b_2$. 

However, $\M{\mu}$ is unknown. We propose to use the $T_X$-invariance property of the Euclidean distance matrix of $X$ to estimate $\V{\mu}$ so that the estimated $\V{\mu}$ can be plugged into equation \eqref{eq:rigid} to estimate $T_X$. Let us first define the Euclidean distance matrix of $X$ as 
\begin{equation*}
\M{D}(X) = \left[ \begin{array}{c c c c c}
0 & d_{\mathbb{X}}(\V{x}_1, \V{x}_2) & d_{\mathbb{X}}(\V{x}_1, \V{x}_3) & \ldots & d_{\mathbb{X}}(\V{x}_1, \V{x}_{m}) \\
d_{\mathbb{X}}(\V{x}_2, \V{x}_1) & 0 & d_{\mathbb{X}}(\V{x}_2, \V{x}_3) & \ldots & d_{\mathbb{X}}(\V{x}_2, \V{x}_{m}) \\
d_{\mathbb{X}}(\V{x}_3, \V{x}_1) & d_{\mathbb{X}}(\V{x}_3, \V{x}_2)  & \vdots & \vdots & \vdots \\
\vdots & \vdots  & \vdots & \vdots & \vdots \\
\end{array} \right],
\end{equation*}
where $d_{\mathbb{X}}(\V{x}_i, \V{x}_j) = ||\V{x}_i - \V{x}_j||_2$. The distance matrix is invariant under a Euclidean rigid body transformation, 
\begin{equation*}
\M{D}(X) = \M{D}(T_X(X)) \mbox{ for } T_X \in \mathbb{SE}(\mathbb{X}). 
\end{equation*}
In addition, the matrix $\M{D}(X)$ contains sufficient information that describes the geometrical features of $X$, because $X$ is uniquely determined from $\M{D}(X)$ up to rotations, reflections and translations by applying the multidimensional scaling to $\M{D}(X)$ \cite[Theorem 1]{lele1993euclidean}. Based on these two properties, a $T_X$-invariant distance between two geometries, $T_X(X)$ and $X_0$, can be defined as the measure of similarity between the two Euclidean distance matrices, $\M{D}(X)$ and $\M{D}(X_0)$. Note that we are defining a reference shape for a collection of geometrically similar objects, so the reference shape preferably have a geometry similar to the objects in the collection as a representative of the collection. Therefore, the Euclidean distance matrices of $T_X(X)$ and $X_0$ should be comparable, i.e., 
\begin{equation*}
d_{\mathbb{X}}(T_X(\V{x}_i), T_X(\V{x}_k)) \approx d_{\mathbb{X}}(\V{x}^{(0)}_j, \V{x}^{(0)}_l) \mbox{ for every } \mu_{ij} = 1, \mu_{kl}=1.
\end{equation*}
Collectively, the equalities are represented by
\begin{equation*}
\M{D}(T_X(X)) \approx \M{\mu} \M{D}(X_0) \M{\mu}^T.
\end{equation*}
Due to the $T_X$-invariance of an Euclidean distance matrix, it also implies 
\begin{equation*}
\M{D}(X)\approx \M{\mu} \M{D}(X_0) \M{\mu}^T.
\end{equation*}
Let $d_{\mathbb{D}}(X, X_0;\V{\mu}) = || \M{D}(X) - \M{\mu} \M{D}(X_0) \M{\mu}^T ||_F$, where $||\cdot||_F$ is the Frobenius norm. We will find $\M{\mu}$ that minimizes the distance,
\begin{equation} \label{eq:D}
\begin{split}
d_{\mathbb{D}}(X, X_0) = \min_{\M{\mu} \in \mathbb{M}_{X,X_0}} d_{\mathbb{D}}(X, X_0;\V{\mu}),
\end{split}
\end{equation}
where $\mathbb{M}_{X,X_0}:= \{(\mu_{ij}) \in \{0, 1\}^{m \times m_0}: \sum_{i=1}^{m} \mu_{ij} \ge 1, \sum_{j=1}^{m_0} \mu_{ij} \ge 1 \}$ defines the range of $\V{\mu}$, and it was defined to make sure that one element in $T_X(X)$ is mapped to at least one element in $X_0$ and vice versa. The algorithm to solve the optimization problem in \eqref{eq:D} can be found in the online supplementary material. Once $\V{\mu}$ is estimated, the estimate can be plugged into equation \eqref{eq:rigid} to estimate the two parameters of $T_X$. It is noteworthy that there is another way to estimate $\V{\mu}$, which finds simultaneously $\V{\mu}$ and $T_X$ by solving
\begin{equation} \label{eq:Drigid}
\begin{split}
\min_{\V{\mu} \in \mathbb{M}_{X,X_0}} \min_{T_X \in \mathbb{SE}(\mathbb{X})}  dist(T_X(X), X_0; \V{\mu}).
 \end{split}
\end{equation}
The optimization has been popularly used for shape matching or two point-set matching \citep{memoli2007use}. The similar formulation was also proposed in statistical shape analysis \citep{rangarajan1997softassign}. The optimization is very complicated \citep{rangarajan1997softassign, green2006bayesian}, because it requires an alternating optimization for $T_X$ and $\V{\mu}$, which often finds local optimality.

Now we explain how the estimation of $T_X$ is applied for an example involving aggregations of different shapes. Suppose that we have $2N$ primary geometric objects from $N$ different aggregation observations. Since the primary objects can have different geometries, we use the similarity measure $d_{\mathbb{D}}(X, X_0)$ to group the $2N$ primary objects by geometric similarities into $K$ shape categories and define a reference shape for each shape category. In this paper, we use the $k$-means clustering with distance $d_{\mathbb{D}}$, where $K$ was chosen using the information criterion, AIC \citep{akaike1992information}. Suppose that $N_k$ geometric objects are grouped to the $k$th shape category, and $X_{n}^{(k)} \subset \mathbb{X}$ denote the $n$th geometric object from the shape category. We choose a cluster representative of the shape category and define it as a reference shape for the shape category. The cluster representative is chosen among $\{X_n^{(k)}; n = 1,..., N_k\}$ so that it minimizes the average distance to the other cluster members. If the cluster representative is $X_r^{k}$, $r$ should satisfy
\begin{equation*}
r = \argmin_{n = 1,\ldots, N_k} \sum_{n' =1}^{N_k} d_{\mathbb{D}}(X_{n'}^{(k)}, X_n^{(k)}).
\end{equation*}
We normalize out the location and orientation of the cluster representative by applying the classical multidimensional scaling to $X_r^{(k)}$. The multidimensional scaling first applies the double centering on $\M{D}(X_r^{(k)})$, subsequently takes the eigen-decomposition on the doubly centered matrix, and finally computes the matrix composed of the eigenvectors scaled by the square roots of the corresponding eigenvalues \citep{lele1993euclidean}. Since the rank of $\M{D}(X_r^{(k)})$ is two, the output matrix of the multidimensional scaling has two columns, and each row vector of the output matrix represents a point coordinate in $\mathbb{R}^2$. Let $\tilde{X}_r^{(k)}$ denote a set of the row vectors in the matrix. It is easy to verify $\M{D}(X_r^{(k)}) = \M{D}(\tilde{X}_r^{(k)})$ so $d_{\mathbb{D}}(X_r^{(k)}, \tilde{X}_r^{(k)}) = 0$. Therefore $\tilde{X}_r^{(k)}$ represents the exactly same geometry as $X_r^{(k)}$. The major axis of $\tilde{X}_r^{(k)}$ is always along the $x$-axis in that the first coordinates of the elements in $\tilde{X}_r^{(k)}$ were generated from the first eigenvector in the multidimensional scaling. Therefore, $\tilde{X}_r^{(k)}$ can be seen as a version of $X_r^{(k)}$ with its orientation normalized. We define $\tilde{X}_r^{(k)}$ as a reference shape for the $k$th shape category. We will present our simulation study in Section \ref{sec:simu} for validating the approaches proposed in this section.

\subsection{Estimation of $\phi_X$, $\phi_Y$ and $\V{c}_{X,Y}$} \label{sec:aggregate}
This section describes how to estimate $\V{c}_{X,Y}$ and the parameters of $\phi_X$ and $\phi_Y$. Let $X \subset \mathbb{X}$ and $Y \subset \mathbb{X}$ denote two primary objects, and let $Z \subset \mathbb{X}$ denote the aggregate of the two primary objects. Suppose that $X$, $Y$ and $Z$ consist of $m_X$, $m_Y$, and $m_Z$ point coordinates respectively,
\begin{equation*}
\begin{split}
&X = \{\V{x}_i \in \mathbb{X}; i = 1,\ldots, m_X\}\\
&Y = \{\V{y}_j \in \mathbb{X}; j = 1,\ldots, m_Y\}\\
&Z = \{\V{z}_k \in \mathbb{X}; k = 1,\ldots, m_Z\}.
\end{split}
\end{equation*}
Since $Z= \phi_X(X) \cup \phi_Y(Y)$, some points in $Z$ correspond to $\phi_X(X)$, and the other points correspond to $\phi_Y(Y)$. Let $\M{\mu}^X = (\mu_{ik}^X)$ define the following measure of correspondences in between elements of the two sets, $\phi_X(X)$ and $Z$, 
\begin{equation*} 
\mu_{ik}^X = 1 \mbox{ if } \phi_{X}(\V{x}_i) \mbox{ corresponds to } \V{z}_k \mbox{ and } 0 \mbox{ otherwise}.
\end{equation*}
Likewise, let $\M{\mu}^Y = (\mu_{jk}^Y)$ denote the elementwise correspondence from $\phi_Y(Y)$ to $Z$. Please note that the $(\M{\mu}^X, \M{\mu}^Y)$ ranges for
\begin{equation*}
\begin{split}
\mathbb{M}_{X,Y;Z} = \{ (\M{\mu}^X, \M{\mu}^Y) : & \sum_{k=1}^{m_Z} \mu_{ik}^X \ge 1 \quad \forall i=1,..,m_X, \\
& \sum_{k=1}^{m_Z}  \mu_{jk}^Y \ge 1  \quad \forall j = 1,...,m_Y, \\
& \sum_{i=1}^{m_X} \mu_{ik}^X + \sum_{j=1}^{m_Y} \mu_{jk}^Y \ge 1  \quad \forall k = 1,...,m_Z \},
\end{split}
\end{equation*}
where the first two inequalities imply that each element in $\phi_X(X)$ and $\phi_X(X)$ corresponds to at least one element in $Z$ and the last inequality implies that each element in $Z$ corresponds to an element in either $\phi_X(X)$ or $\phi_Y(Y)$. When $(\M{\mu}^X, \M{\mu}^Y)$ are known, the two rigid body transformations $\phi_X \in \mathbb{SE}(\mathbb{X})$ and $\phi_Y \in \mathbb{SE}(\mathbb{X})$ can be estimated by solving
\begin{equation*}
\begin{split}
(\phi_X^*, \phi_Y^*) = \argmin_{\phi_X, \phi_Y \in \mathbb{SE}(\mathbb{X})} \sum_{i=1}^{m_X}\sum_{k=1}^{m_Z} \mu_{i,k}^X \left|\left|\phi_X(\V{x}_i) - \V{z}_k\right|\right|^2 +  \sum_{j=1}^{m_Y}\sum_{k=1}^{m_Y} \mu_{jk}^Y \left|\left|\phi_Y(\V{y}_j) - \V{z}_k\right|\right|^2.  
\end{split}
\end{equation*}
Let $\V{c}^*_{\phi_X}$ and $\theta^*_{\phi_X}$ denote the translation vectors and the rotation angle of $\phi_X^*$, and let $\V{c}^*_{\phi_Y}$ and $\theta^*_{\phi_Y}$ denote those of $\phi_Y^*$. The parameter values can be achieved using the first order necessary condition as follows,
\begin{equation} \label{eq:rigid2}
\begin{split}
& \V{c}^*_{\phi_X} = \frac{\sum_{i=1}^{m_X} \sum_{k=1}^{m_Z} \mu_{ik}^X (\V{x}_i - \V{z}_k)}{\sum_{i=1}^{m_X} \sum_{k=1}^{m_Z} \mu_{ik}^X}, \theta^*_{\phi_X} = \arctan\left(\frac{\sum_{i=1}^{m_X} \sum_{k=1}^{m_Z}  \mu_{ik}^X (\V{z}_k \times \V{x}_i)}{\sum_{i=1}^{m_X} \sum_{k=1}^{m_Z} \mu_{ik}^X (\V{z}_k \cdot \V{x}_i) }\right)\\
& \V{c}^*_{\phi_Y} = \frac{\sum_{j=1}^{m_Y} \sum_{k=1}^{m_Z} \mu_{jk}^Y (\V{y}_i - \V{z}_k)}{\sum_{j=1}^{m_Y} \sum_{k=1}^{m_Z} \mu_{jk}^Y}, \theta^*_{\phi_Y} = \arctan\left(\frac{\sum_{j=1}^{m_Y} \sum_{k=1}^{m_Z}  \mu_{jk}^Y (\V{z}_k \times \V{y}_j)}{\sum_{j=1}^{m_Y} \sum_{k=1}^{m_Z} \mu_{jk}^Y (\V{z}_k \cdot \V{y}_j) }\right).
\end{split}
\end{equation} 
Since $(\M{\mu}^X, \M{\mu}^Y)$ are unknown, similar to what we did in the previous section, we use the Euclidean distance matrices of $X$, $Y$ and $Z$ to estimate $(\M{\mu}^X, \M{\mu}^Y)$, 
\begin{equation} \label{eq:d2}
\begin{split}
\min_{(\M{\mu}^X, \M{\mu}^Y) \in \mathbb{M}_{X,Y;Z}} d_{\mathbb{D}}(X, Z;\M{\mu}_X) + d_{\mathbb{D}}(Y, Z;\M{\mu}_Y).
\end{split}
\end{equation}
The algorithm to solve the optimization problem can be found in the online supplementary material. The optimal solution provides the point-to-point correspondence $(\M{\mu}^X, \M{\mu}^Y)$. By plugging $(\M{\mu}^X, \M{\mu}^Y)$ in the expression \eqref{eq:rigid2}, the $\phi_X$ and $\phi_Y$ can be estimated. 

In addition, the aggregation center of $Z$ can be estimated with $(\M{\mu}^X, \M{\mu}^Y)$ by first finding the subset of $Z$ that corresponds to both $X$ and $Y$, 
\begin{equation*} 
C_{X,Y} = \{ \V{z}_k \in Z: \mu_{ik}^X = 1 \mbox{ and } \mu_{jk}^Y = 1\},
\end{equation*}
and then estimating the mass center of $C_{X,Y}$,
\begin{equation} \label{eq:agg_center}
\V{c}_{X,Y} = \frac{\sum_{\V{z}_k \in C_{X,Y}} \V{z}_k}{|C_{X,Y}|},
\end{equation}
where $|\cdot|$ is the number of elements in a set. A combination of this result with the estimation of the $\phi_X$ and $\phi_Y$ is used to evaluate $\phi_X^{-1}(\V{c}_{X,Y})$ and $\phi_Y^{-1}(\V{c}_{X,Y})$.

\subsection{Simulation study} \label{sec:simu}
We performed a simulation study to numerically validate the proposal approaches described in the previous subsections. In this simulation study, we examined our approach for datasets emulating aggregations of ellipses. An ellipse was chosen because it is the simplest object that has directionality. Certainly, there are infinitely many types of other shapes that have directionality, but it is impossible to test the proposed approach numerically for all those cases. The shapes of aggregating objects are often dependent on the areas of application. This section provides at least a general guideline for practitioners to implement the similar type of numerical studies with other shapes prevailing in their applications. This limited validation does not mean that the proposed approach works only for ellipses. 

Simulation inputs were shape factors of primary objects, the variations of the shape factors, and the levels of observation noises. Since we restricted the shapes of primary particles to ellipses, the shape factor is characterized by the major axis length and the minor axis length. We followed the following generative procedure to simulate a set of 50 aggregation cases,
\begin{description}
\item[Inputs:] Let $X$ and $Y$ denote two ellipses to be aggregated. Let $a_X$ and $b_X$ denote the major axis length and the minor axis length of an ellipse $X$. Let $a_Y$ and $b_Y$ denote the major axis length and the minor axis length of an ellipse $Y$.  The inputs of the simulation are the variations of the shape factors,\\
$\nu_{a,X}$: the logarithm of the mean of $a_X$,  \\
$\nu_{a,Y}$: the logarithm of the mean of $a_Y$, \\
$\nu_{b,X}$: the logarithm of the mean of $b_X$, \\
 $\nu_{b,Y}$: the logarithm of  the mean of $b_Y$, \\
$\sigma^2$: shape variations, and $\sigma_e^2$: noise variance. 
\item[Step 1.] \textbf{Simulate $X$:} Sample $\log(a_X) \sim \mathcal{N}(\nu_{a,X}, \sigma^2)$ and $\log(b_X) \sim \mathcal{N}(\nu_{b,X}, \sigma^2)$. Generate a noisy image of an ellipse,  $\tilde{X} = \left\{(x_1,x_2) \in \mathbb{X}; \frac{x_1^2}{a_X^2} + \frac{x_2^2}{b_X^2} \le 1 + \epsilon(|\frac{x_2}{x_1}|) \right\}$, where $\epsilon(|\frac{x_2}{x_1}|) \sim \mathcal{N}(0, \sigma_{e}^2)$ is a random process depending on $|\frac{x_2}{x_1}|$. Let $T_X$ denote a random Euclidean rigid body transformation with a translation vector $\V{c}_{T_X} \sim \Unif([0, H] \times [0, W])$ and a rotation angle $\theta_{T_X} \sim \Unif([0, \pi/2])$. The noisy image $\tilde{X}$ is transformed to $T_X^{-1}(\tilde{X})$, which serves $X$. 
\item[Step 2.] \textbf{Simulate $Y$:}  Sample $\log(a_Y) \sim \mathcal{N}(\nu_{a,Y}, \sigma^2)$ and $\log(b_Y) \sim \mathcal{N}(\nu_{b,Y}, \sigma^2)$. Generate a noisy image of an ellipse,  $\tilde{Y} = \left\{(x_1,x_2) \in \mathbb{X}; \frac{x_1^2}{a_Y^2} + \frac{x_2^2}{b_Y^2} \le 1 + \epsilon(|\frac{x_2}{x_1}|) \right\}$, where $\epsilon(|\frac{x_2}{x_1}|) \sim \mathcal{N}(0, \sigma_{e}^2)$ is a random process depending on $|\frac{x_2}{x_1}|$. Let $T_Y$ denote a random Euclidean rigid body transformation with a translation vector $\V{c}_{T_Y} \sim \Unif([0, H] \times [0, W])$ and a rotation angle $\theta_{T_Y} \sim \Unif([0, \pi/2])$. The noisy image $\tilde{Y}$ is transformed to $T_Y^{-1}(\tilde{Y})$, which serves $Y$. 
\item[Step 3.] \textbf{Simulate $Z$:}  Let $\phi_X$ denote the Euclidean rigid body transformation with a translation vector $\V{c}_{\phi_X}$ and a rotation angle $\theta_{\phi_X}$. Sample $\V{c}_{\phi_X} \sim \Unif(X)$ and $\theta_X=\pi$-angle$(\V{c}_{\phi_X} + \V{c}_{T_X})$. Let $\phi_Y$ denote the Euclidean rigid body transformation with a translation vector $\V{c}_{\phi_Y}$ and a rotation angle $\theta_{\phi_Y}$. Sample $\V{c}_{\phi_Y} \sim \Unif(Y)$ and $\theta_{\phi_Y}=-angle(\V{c}_{\phi_Y} + \V{c}_{T_Y})$. Let $Z = \phi_X(X) \bigcup \phi_Y(Y)$. 
\item[Step 4.] Repeat Steps 1 through 3 for 50 times. 
\end{description}
We fixed $\nu_{b,X}=\log(5)$ while $\nu_{a,X}$ was varied to $\exp(\nu_{a,X}) = r_X \exp(\nu_{b,X})$, where $r_X$ represents the ratio of the mean major axis length and the mean minor axis length. The choice of $\nu_{b,X}$ is not critical at all, which just determines the overall expected size of a primary object $X$ and is nothing related to its shape factor and directionality. The choice is just one of arbitrary choices. Similarly, the choices of $\nu_{b,Y}=\log(5)$ and $\nu_{a,Y} = \log(r_Y) + \nu_{b,Y}$ are also arbitrary. We fixed $\sigma^2 = 0.03^2$, which makes $\exp(\nu_{b,X})$ or $\exp(\nu_{b,Y})$ approximately range for $[4.5, 5.5]$. We also fixed $\sigma_e^2 = 0.1^2$, which makes $1+\epsilon(|\frac{x_2}{x_1}|)$ approximately range for $[0.97, 1.03]$. We tried six different combinations of $r_X  \in \{1.1, 1.4, 2.2\}$ and $r_Y \in \{1.1, 1.4, 2.2\}$ to simulate simulation cases involving different shape factors. For each of the combinations, we performed 50 replicated experiments, and each of the replicated experiments has 50 aggregation cases.

We applied the methods proposed in Sections \ref{sec:align} and \ref{sec:aggregate} to the simulated datasets to estimate $T_X$, $T_Y$, $\phi_X$ and $\phi_Y$. Note that the $T_X$ is parameterized by two parameters $\V{c}_{T_X}$ and $\theta_{T_X}$, $T_Y$ by $\V{c}_{T_Y}$ and $\theta_{T_Y}$, $\phi_X$ by $\V{c}_{\phi_X}$ and $\theta_{\phi_X}$, and $\phi_Y$ by $\V{c}_{\phi_Y}$ and $\theta_{\phi_Y}$. The estimated parameters are denoted by $\V{c}_{T_X}^*$, $\theta_{T_X}^*$, $\V{c}_{T_Y}^*$, $\theta_{T_Y}^*$, $\V{c}_{\phi_X}^*$, $\theta_{\phi_X}^*$, $\V{c}_{\phi_Y}^*$ and $\theta_{\phi_Y}^*$. For each combination of $r_X$ and $r_Y$, we evaluated the differences of the estimated parameter values and the corresponding simulation inputs over 50 replicated experiments and 50 aggregation cases per experiment. For each translation vector estimate $\V{c}^*$, we took the square difference, $(\V{c} - \V{c}^*)^T (\V{c} - \V{c}^*)$. For each rotation angle estimate $\theta^*$, we took the angular difference, $1-\cos(\theta - \theta^*)$, after some angular normalization steps to compensate for geometric symmetries of ellipses; we will discuss this particular issues in Section \ref{sec:stat}. Table \ref{tbl:simul} summarizes the average differences and the standard deviations of the estimates. For higher $r_X$ (or $r_Y$), the estimation accuracy for $T_X$ (or $T_Y$) increases. Note that with a higher $r_X$ implies a clearer directionality of a primary object. The simulation outcomes explain that a clearer directionality of primary objects would help to align them and estimate $T_X$ accurately. When $r_X$ is below 1.4, the estimation accuracy degrades significantly. We do not suggest to apply the proposed approach for analyzing the aggregations of ellipses with $r_X$ less than $1.4$. For our motivating example, primary nanoparticles with $r_X \ge 1.4$ accounted for 62\% of all primary nanoparticles observed (228 out of 368). Therefore, our approach is applicable for a majority of the cases. In addition, ellipses with $r_X < 1.4$ are very close to circles, for which the spatial directions are not clearly defined. On the other hand, the estimation accuracy of $\phi_X$ or $\phi_Y$ did not depend much on $r_X$ or $r_Y$.  The variations of the estimates were very small when both of $r_X$ and $r_Y$ are greater than or equal to 1.4.

\begin{table}
\centering
\begin{tabular}{|c|c|c|c|c|c|c|c|c|}
\hline
Setting      & \multicolumn{8}{c|}{Parameters} \\ \cline{2-9}
$(r_X, r_Y)$ & $\V{c}_{T_X}^*$ & $\theta_{T_X}^*$ & $\V{c}_{T_Y}^*$ & $\theta_{T_Y}^*$ & $\V{c}_{\phi_X}^*$ & $\theta_{\phi_X}^*$ & $\V{c}_{\phi_Y}^*$ & $\theta_{\phi_Y}^*$ \\
\hline
(2.2, 2.2)   & 0.0190 & 0.0007 & 0.0227 & 0.0010 & 0.0368 & 0.0002 & 0.0304 & 0.0001 \\
             &(0.0028)&(0.0002)&(0.0045)&(0.0002)&(0.0060)&(0.00003)&(0.0044)&(0.00001)\\ \hline
(2.2, 1.4)   & 0.0220 & 0.0003 & 0.0168 & 0.0058 & 0.0342 & 0.0001 & 0.0332 & 0.0002 \\
             &(0.0036)&(0.0001)&(0.0023)&(0.0015)&(0.0050)&(0.00001)&(0.0062)&(0.00007) \\ \hline
(2.2, 1.1)   & 0.0293 & 0.0010 & 0.0152 & 0.0744 & 0.0466 & 0.0001 & 0.0288 & 0.0003 \\
             &(0.0038)&(0.0002)&(0.0020)&(0.0157)&(0.0054)&(0.00002)&(0.0064)&(0.00009) \\ \hline
(1.4, 1.4)   & 0.0174 & 0.0064 & 0.0246 & 0.0034 & 0.0291 & 0.0002 & 0.0312 & 0.0001 \\
             &(0.0025)&(0.0009)&(0.0037)&(0.0012)&(0.0044)&(0.00003)&(0.0050)&(0.00002) \\ \hline
(1.4, 1.1)   & 0.0252 & 0.0046 & 0.0155 & 0.0520 & 0.0304 & 0.0001 & 0.0247 & 0.0002 \\
             &(0.0040)&(0.0009)&(0.0022)&(0.0135)&(0.0043)&(0.00002)&(0.0036)&(0.00005) \\ \hline
(1.1, 1.1)   & 0.0204 & 0.0587 & 0.0161 & 0.0966 & 0.0280 & 0.0001 & 0.0218 & 0.00001 \\
             &(0.0025)&(0.0128)&(0.0018)&(0.0218)&(0.0032)&(0.00002)&(0.0027)&(0.00002) \\
\hline
\end{tabular}
\caption{Accuracy of parameter estimation for $T_X$, $T_Y$, $\phi_X$ and $\phi_Y$. We evaluated the differences of the estimated parameter values and the corresponding simulation inputs over 50 replicated experiments and 50 aggregation cases per experiment. The first number of each cell is the mean squared difference for $\V{c}^*$ or the average of the angular difference for $\theta^*$. The second number surrounded by two round brackets is the standard deviation of an estimate over replicated experiments.} \label{tbl:simul}
\end{table}

\section{STATISTICAL ANALYSIS OF AGGREGATION} \label{sec:stat}
The major scientific questions that we want to answer were (1) whether there are preferential orientations of primary objects when they aggregate, and (2) if so, what the orientations are. In this section, we present a statistical analysis to answer those questions. 

Suppose that we have $N$ aggregation observations,
\begin{equation*}
\{(X_n,Y_n,Z_n); n=1,....N \},
\end{equation*}
where $X_n$ and $Y_n$ are the simply connected subsets of $\mathbb{X}$ that represents two primary geometric objects for the $n$th observation, and $Z_n$ is the simply connected subset of $\mathbb{X}$ that represents the corresponding aggregate. As described in Section \ref{sec:align}, the $2N$ primary objects are grouped into $K$ shape categories based on their geometric similarities, and for each shape category, we identified a reference shape and had all primary objects in the category aligned to the reference shape to define the standard coordinate systems for the primary objects. 

Some shape categories may have geometrical symmetries around their major axes and minor axes, e.g., a rod and an ellipse. The major axis of a geometric object $X_n$ is defined by the first principal loading vector of the coordinates in $X_n$, and the minor axis is the unit vector perpendicular to the major axis. Note that with the alignment described in Section \ref{sec:align}, the major axis of a primary object is along the $x$-axis, and the minor axis is along the $y$-axis. For the primary objects belonging to a shape category symmetric around the major and minor axis, the following orientation angles of the primary objects are indistinguishable due to the geometrical symmetry,
\begin{equation} \label{eq:equiv}
\theta \equiv -\theta \equiv \pi - \theta \equiv -\pi + \theta \mbox{ for } \theta \in [0, \pi/2].
\end{equation}
Therefore, for a symmetric shape category, we normalize orientation $\theta$ to
\begin{equation} \label{eq:nang}
\tilde{\theta} =
\begin{cases}
|\theta|  & \mbox{ if } |\theta| \le \pi/2, \\
\pi-|\theta| & \mbox{otherwise,}
\end{cases}
\end{equation}
which is basically one of the $\theta$'s equivalent forms in the first quadrant $[0, \pi/2]$. 

We perform statistical inference on an unnormalized angle $\theta$ for a non-symmetric shape category and on an normalized angle $\tilde{\theta}$ for a symmetric shape category. The probability distribution of $\theta$ for a non-symmetric case can be modeled as a von Mises distribution, which is popularly used to describe a unimodal probability density of angular data \citep{mardia2012mixtures}. The statistical inferences on the distribution model have been well studied in circular statistics \citep{fisher1995statistical}; therefore, we will not reiterate them in this paper. This section focuses on statistical analysis of $\tilde{\theta}$ for symmetric cases.

For a symmetric shape category, the equivalence \eqref{eq:equiv} holds in $\theta$, and the probability density function of $\theta$ should have the following symmetries,
\begin{equation} \label{eq:sym}
f(\theta) = f(-\theta) = f(-\pi + \theta) = f(\pi - \theta).
\end{equation}
Therefore, if $f$ has a mode at $\gamma \in [0, \pi/2]$, it also has the modes at $-\gamma$, $-\pi + \gamma$ and $\pi - \gamma$. A von-Mises distribution is popularly used to describe a unimodal probability density of angular data \citep{mardia2012mixtures}. We take a mixture of four von Mises distributions with equal weights to represent the four modes caused by the four-way symmetry, 
\begin{equation*}
\begin{split}
f(\theta; \gamma, \kappa) = &  \frac{1}{8\pi I_0(\kappa)} \exp\{ \kappa \cos(\theta - \gamma) \}  + \frac{1}{8\pi I_0(\kappa)}\exp\{ \kappa \cos(\theta + \pi- \gamma)\} \\
                         & + \frac{1}{8\pi I_0(\kappa)} \exp\{ \kappa \cos(\theta + \gamma) \} + \frac{1}{8\pi I_0(\kappa)}\exp\{ \kappa \cos(\theta - \pi + \gamma) \}\\
                       = &  \frac{1}{8\pi I_0(\kappa)} \exp\{ \kappa \cos(\theta - \gamma) \}  + \frac{1}{8\pi I_0(\kappa)}\exp\{ -\kappa \cos(\theta - \gamma)\} \\
                        & + \frac{1}{8\pi I_0(\kappa)} \exp\{ \kappa \cos(\theta + \gamma) \} + \frac{1}{8\pi I_0(\kappa)}\exp\{ -\kappa \cos(\theta + \gamma) \}\\
                      = & \frac{1}{4\pi I_0(\kappa)} \cosh(\kappa \cos(\theta - \gamma)) + \frac{1}{4\pi I_0(\kappa)} \cosh(\kappa \cos(\theta + \gamma)) \\
                      = & \frac{1}{2\pi I_0(\kappa)} \cosh(\kappa \cos(\gamma) \cos(\theta)) \cosh(\kappa \sin(\gamma) \sin(\theta)),
\end{split}
\end{equation*}
where $\cosh(\cdot)$ is a hyperbolic cosine function, and $\gamma \in [0, \pi/2]$. One can easily check that the density function satisfies the symmetry \eqref{eq:sym} as desired. Note that the normalization \eqref{eq:nang} applies for mirroring $\theta$ onto the first quadrant $[0, \pi/2]$, and $f$ has the same density for all quadrants. Therefore, the density function of the normalized angle $\tilde{\theta}$ is simply four times of $f$, 
\begin{equation}
g(\tilde{\theta}; \gamma, \kappa) = \frac{2}{\pi I_0(\kappa)} \cosh(\kappa \cos(\gamma) \cos(\tilde{\theta})) \cosh(\kappa \sin(\gamma) \sin(\tilde{\theta})),
\end{equation}
where $\gamma, \tilde{\theta} \in [0, \pi/2]$. One can show $\int_0^{\pi/2} g(\tilde{\theta}; \gamma, \kappa) = 1$, so it is a valid probability density function. The two parameters $\gamma$ and $\kappa$ can be estimated by the maximum likelihood estimation described in Section \ref{sec:mle}, and the goodness-of-fit test for the estimated parameters can be performed by the method described in Section \ref{sec:fit}. Sections \ref{sec:uniftest} and \ref{sec:meantest} describes the statistical hypotheses testing problems to test the two hypotheses that we mentioned in the beginning of this section. 

\subsection{Maximum Likelihood Estimation} \label{sec:mle}
We present a numerical procedure to compute maximum likelihood estimates of $\gamma$ and $\kappa$ for $g(\tilde{\theta}; \gamma, \kappa)$, given a random sample $\{\tilde{\theta}_1, \ldots, \tilde{\theta}_N\}$ from the density. The log likelihood function is 
\begin{equation} \label{eq:like}
L_N(\gamma, \kappa) = \sum_{n=1}^N \log (\cosh(\kappa \cos(\gamma) \cos(\tilde{\theta}_n))) + \log (\cosh(\kappa \sin(\gamma) \sin(\tilde{\theta}_n))) - N\log (I_0(\kappa)). 
\end{equation}
The first order necessary condition, $\frac{\partial L_N}{\partial \gamma} = 0$ and $\frac{\partial L_N}{\partial \kappa} = 0$, does not give a closed form expression for $\gamma$ and $\kappa$. The two parameters $\gamma$ and $\kappa$ can be numerically optimized by the Newton-Raphson algorithm using the first order derivatives and the second order derivatives of the log likelihood function. The expressions for the first and second order derivatives can be found in the online supplementary material. The optimization algorithm starts with initial guesses on the parameter values and iteratively change the values toward the increasing direction of the likelihood \eqref{eq:like}. A possible initial guess for $\gamma$ can be the sample angular mean $s_{\gamma}$, and a possible initial guess for $\kappa$ can be $s_{\kappa}$, the unbiased estimator of $\frac{I_1(\kappa)}{I_0(\kappa)}$, 
\begin{equation*}
s_{\gamma} = \arctan\left( \frac{\bar{s}}{\bar{c}} \right) \mbox{ and } \frac{I_1(s_{\kappa})}{I_0(s_{\kappa})} = \frac{N}{N-1} \bar{c}^2 + \bar{s}^2 - \frac{1}{N-1},
\end{equation*}
where $\bar{s} = \frac{1}{N}\sum_{n=1}^N \sin(\tilde{\theta}_n)$ and $\bar{c} = \frac{1}{N}\sum_{n=1}^N \cos(\tilde{\theta}_n)$. Since the Newton-Raphson algorithm may find a local optimum, we ran the algorithm with different initial guesses ranging $\gamma \in \{s_{\gamma}-0.1, s_{\gamma}, s_{\gamma}+0.1\}$ and $\kappa \in \{s_{\kappa}-1, s_{\kappa}, s_{\kappa}+1\}$. Among the trials, we chose one that gave the highest likelihood value at the end of the algorithm. 

We evaluated the bias and variance of the maximum likelihood estimates resulting from the numerical procedure using three simulation cases. We first drew a random sample of size 1000 from $g(\tilde{\theta}; \gamma, \kappa)$ with $\gamma$ and $\kappa$ specified in Table \ref{tbl:mle} and used the random sample to estimate $\gamma$ and $\kappa$ as described in this section. The estimates $\hat{\gamma}$ and $\hat{\kappa}$ were compared with the values of $\gamma$ and $\kappa$ used as simulation inputs, and their differences were calculated for the biases of the two estimates. The random sampling followed by the maximum likelihood estimation was repeated 100 times, and the biases for the replicated experiments were averaged, and the variances of the estimates over 100 replicated experiments were evaluated. Table \ref{tbl:mle} summarizes the outcomes. The biases and variances of $\hat{\gamma}$ were very small, and the biases of $\hat{\kappa}$ are a little higher but still close to zero. 


\begin{table}
\centering
\begin{tabular}{|l|c|c|c|c|c|c|}
\hline
\multirow{2}{*}{Simulation Inputs} & \multicolumn{2}{|c}{$\gamma = \pi/6, \kappa = 10$} &  \multicolumn{2}{|c|}{$\gamma = \pi/4, \kappa = 10$} & \multicolumn{2}{|c|}{$\gamma = \pi/6, \kappa = 5$} \\\cline{2-7}
                   & $\hat{\gamma}$ & $\hat{\kappa}$ & $\hat{\gamma}$ & $\hat{\kappa}$ &$\hat{\gamma}$ & $\hat{\kappa}$ \\ 
\hline
Bias & 0.0017 & 0.0488 &  0.0012 & 0.0274 & 0.0032 & 0.0503 \\
Variance & 0.0001 & 0.0036 & 0.0001 &  0.0032 & 0.0005 & 0.0057 \\
\hline
\end{tabular}
\caption{Biases and standard deviations of the maximum likelihood estimates $\hat{\gamma}$ and $\hat{\kappa}$. A value in each cell is the average value over 100 replicated runs.}
\label{tbl:mle}
\end{table}

\subsection{Goodness-of-Fit Test}\label{sec:fit}
We use the Kolmogorov-Smirnov test \citep{arnold2011nonparametric} to test the goodness-of-fit of $g(\tilde{\theta}; \hat{\gamma}, \hat{\kappa})$ to a random sample $\{\tilde{\theta}_1, \ldots, \tilde{\theta}_N\}$. Let $G(\tilde{\theta})$ denote the cumulative distribution function that corresponds to $g(\tilde{\theta}; \hat{\mu}, \hat{\kappa})$, and let $G_n(\tilde{\theta})$ denote the empirical cumulative distribution function, 
\begin{equation*}
G_n(\tilde{\theta}) = \frac{1}{N} \sum_{n=1}^N I_{[-\infty, \tilde{\theta}]} (\tilde{\theta}_n).
\end{equation*}
The test statistic for the Kolmogorov-Smirnov test is the difference in between the two cumulative distribution functions defined as follows, 
\begin{equation*}
T_N = \sqrt{n} \sup_{\tilde{\theta}} |G(\tilde{\theta}) - G_n(\tilde{\theta})|.
\end{equation*}
If the test statistic is below a critical value, the fit of $G$ to $G_n$ is good. The critical value is determined so that the type-I error is $\alpha$, which is denoted by $t_{\alpha, N}$.  The critical value can be achieved by the following Monte Carlo simulation, 
\begin{description}
\item[Step 1.] Take a random sample of size $N$ from $g(\tilde{\theta}; \hat{\gamma}, \hat{\kappa})$, and get the empirical cumulative distribution function $G_n$ for the random sample. 
\item[Step 2.] Compute $T_N$. 
\item[Step 3.] Repeat Step 1 and Step 2 many times, which results in a number of $T_N$ values. The critical value of the test statistic with type-I error $\alpha$ is the $1-\alpha$ quantile of the resulting $T_N$ values.
\end{description}

\subsection{Testing the Uniformity of Distribution} \label{sec:uniftest}
The first hypothesis to test is whether there is a preferential orientation of a primary object in its aggregate. It is related to testing whether $g(\tilde{\theta}; \gamma, \kappa)$ is uniform, since more uniformity implies less preferential orientation. The uniformity of the density function $g(\tilde{\theta}; \gamma, \kappa)$ is determined by its parameter $\kappa$. Note as the parameter value decreases, the density function $g(\tilde{\theta}; \gamma, \kappa)$ becomes closer to an angular uniform distribution and becomes perfectly uniform with $\kappa = 0$ and nearly uniform with $\kappa \le 0.5$. Therefore, we formulate the uniformity testing as follows,
\begin{equation*}
\begin{split}
\mbox{H}_0: & \kappa \le 0.5 \\
\mbox{H}_1: & \kappa > 0.5.
\end{split}
\end{equation*}
We can test the hypothesis based on a general likelihood ratio test with a random sample from $g(\tilde{\theta}; \gamma, \kappa)$. Suppose that $\{\tilde{\theta}_1, \ldots, \tilde{\theta}_N\}$ is the random sample. Using the likelihood function \eqref{eq:like}, we can define the likelihood ratio test statistic for testing H$_0$ versus H$_1$,
\begin{equation*}
R_{\kappa} = \max_{\kappa > 0.5} L_{N}(\gamma, \kappa)- \max_{\kappa \le 0.5} L_{N}(\gamma, \kappa).
\end{equation*}
Evaluating the test statistic involves evaluating two maximum likelihoods under different linear constraints on $\kappa$, which can be solved easily using the Newton Raphson algorithm as we described in Section \ref{sec:mle}. When the test statistic is above a critical value, we reject H$_0$. The critical value of the test statistic with type-I error $\alpha$ can be easily determined using the following Monte Carlo simulation,
\begin{description}
\item[Step 1.] Sample $\kappa \sim \Unif([0, 0.5])$ and $\gamma \sim \Unif([0, \pi/2])$. 
\item[Step 2.] Take a random sample of size $N$ from $g(\tilde{\theta}; \gamma, \kappa)$, and evaluate $R_{\kappa}$ for the random sample. 
\item[Step 3.] Repeat Steps 1 and 2 many times, which results in a number of $R_{\kappa}$ values. The critical value of the test statistic with type-I error $\alpha$ is the $1-\alpha$ quantile of the resulting $R_{\kappa}$ values.
\end{description}  

\subsection{Testing the Mean Orientation} \label{sec:meantest}
The second hypothesis to test is whether the mean orientation of a primary object in its aggregate is $\gamma_0$. When the orientation follows the probability density $g(\tilde{\theta}; \gamma, \kappa)$, this test can be formulated as testing whether $\gamma = \gamma_0$ or not. We can test the hypothesis based on a general likelihood ratio test with a random sample from $g(\tilde{\theta}; \gamma, \kappa)$. Suppose that $\{\tilde{\theta}_1, \ldots, \tilde{\theta}_N\}$ is the random sample. The likelihood ratio test statistic is
\begin{equation*}
R_{\gamma}  = \max_{\gamma, \kappa} L_{N}(\gamma, \kappa)- \max_{\gamma=\gamma_0,\kappa} L_{N}(\gamma, \kappa).
\end{equation*}
Evaluating the test statistic involves evaluating two maximum likelihoods, one with no constraint and another with a linear constraint on $\gamma$, which can be solved easily using the Newton Raphson algorithm as we described in Section \ref{sec:mle}. When the test statistic is below a critical value, it implies that there is no significant evidence to refuse $\gamma = \gamma_0$. The critical value of the test statistic with type-I error $\alpha$ can be easily determined using the following Monte Carlo simulation,
\begin{description}
\item[Step 1.] Sample $\kappa \sim \Unif([0, 30])$ and $\gamma = \gamma_0$. 
\item[Step 2.] Take a random sample of size $N$ from $g(\tilde{\theta}; \gamma, \kappa)$, and evaluate $R_{\gamma}$ for the random sample. 
\item[Step 3.] Repeat Steps 1 and 2 many times, which results in a number of $R_{\gamma}$ values. The critical value of the test statistic with type-I error $\alpha$ is the $1-\alpha$ quantile of the resulting $R_{\gamma}$ values.
\end{description}  

\section{APPLICATION TO NANOPARTICLE AGGREGATION} \label{sec:exp}
\begin{figure}[p]
\centering
\subfigure[Shape Category 1: Rod]{%
\includegraphics[width=0.5\textwidth]{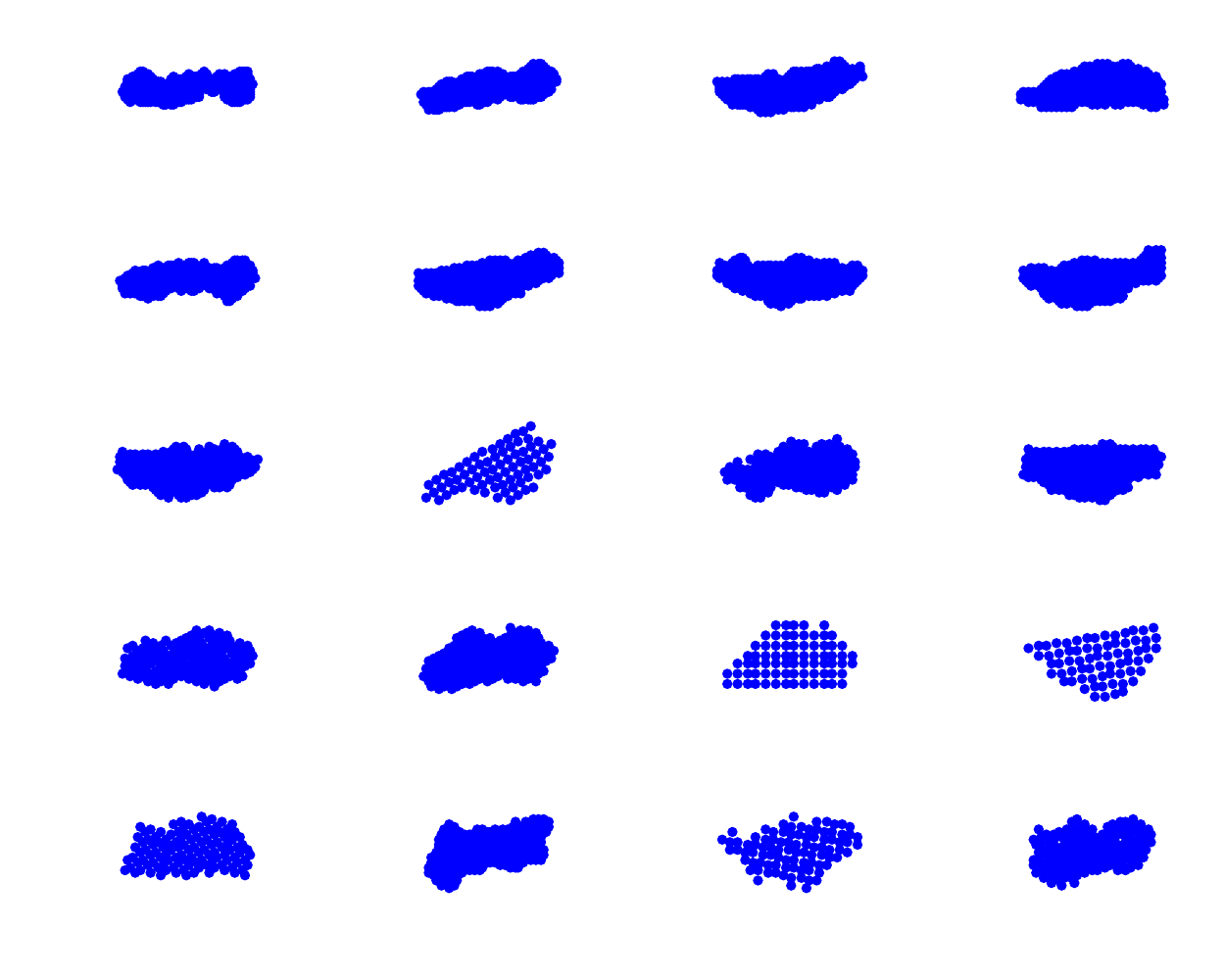}
\label{fig:subfigure1}}
\subfigure[Shape Category 2: Ellipse]{%
\includegraphics[width=0.5\textwidth]{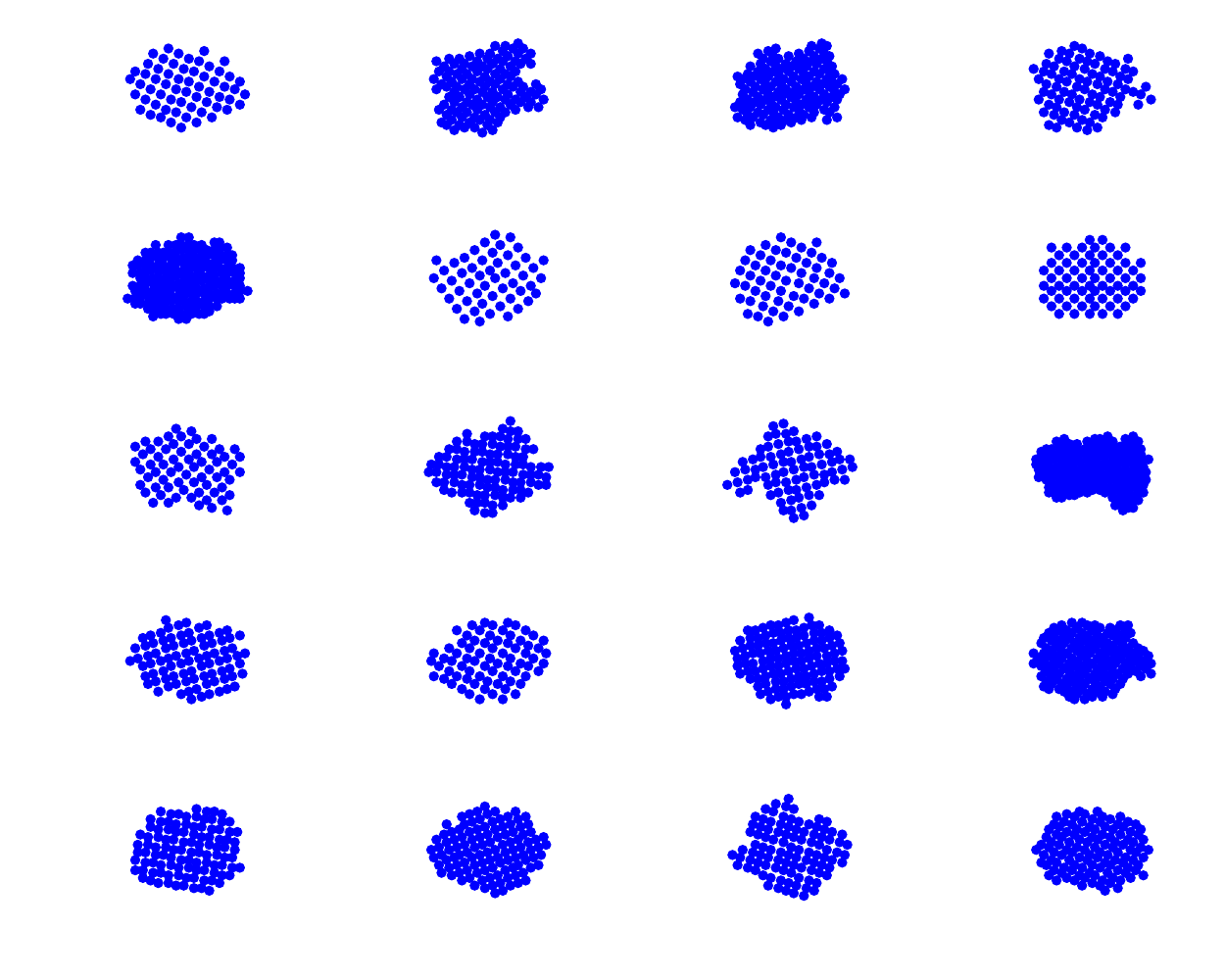}
\label{fig:subfigure2}}
\subfigure[Shape Category 3: NearSphere]{%
\includegraphics[width=0.5\textwidth]{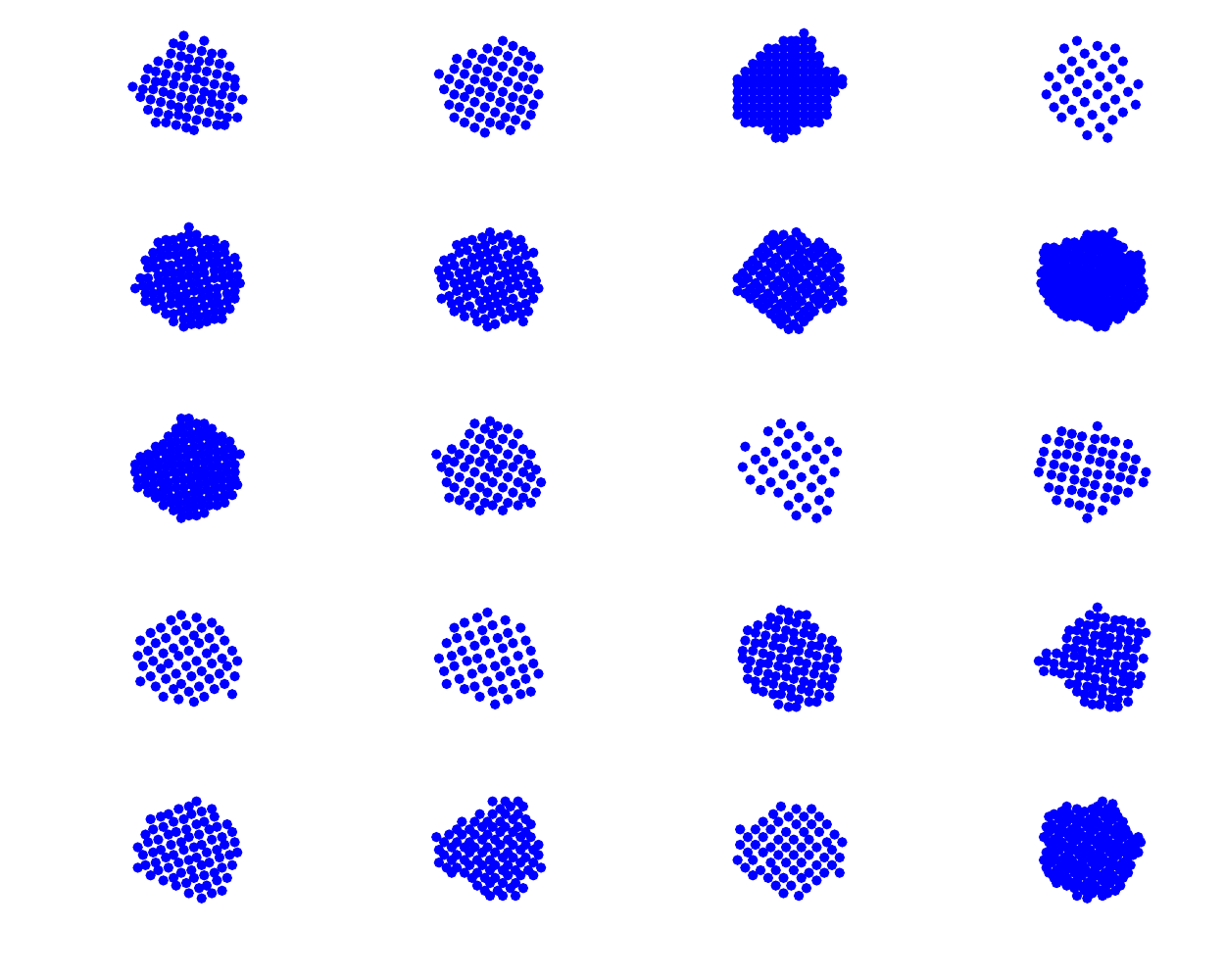}
\label{fig:subfigure3}}
\caption{Alignment outcomes for three shape categories}
\label{fig:objects}
\end{figure}

The motivating example described in Section \ref{sec:dataset} provided 184 aggregation observations for nanoparticles, i.e., $N=184$. The method proposed in Section \ref{sec:align} was applied to group the $2N$ primary objects into $K$ shape categories by their geometric similarities; $K=3$ was chosen by the AIC. For each shape category, we identified a reference shape, and the primary objects in the category were aligned to the reference shape, based on \eqref{eq:rigid}. Figure \ref{fig:objects} illustrates the images of the primary particles after the alignment. Notably, the major axes of the primary particles were aligned to the horizontal line (i.e. $x$-axis), which indicates that the alignment task worked well. Apparently those three shape categories are distinct in terms of an aspect ratio, which is defined as the ratio of the major axis length and the minor axis length of a shape. The mean aspect ratios are 1.99 for the first category, 1.40 for the second, and 1.22 for the last category. Based on the typical appearances of nanoparticles, we named shape category 1 as 'Rod' ($k=1$, 82 objects), shape category 2 as 'Ellipse' ($k=2$, 146 objects), and shape category 3 as 'NearSphere' ($k=3$, 140 objects).

The $N$ aggregation observations can be classified into six groups, depending on the shape categories of the primary objects involved in the aggregations, Rod-Rod (12 cases), Rod-Ellipse (26 cases), Rod-NearSphere (32 cases), Ellipse-Ellipse (33 cases), Ellipse-NearSphere (54 cases), and NearSphere-NearSphere (27 cases).  We achieved the orientation angles of primary nanoparticles normalized to $[0, \pi/2]$ as described in Section \ref{sec:stat}, 
\begin{equation*}
\{(\tilde{\theta}_X^{(n)}, \tilde{\theta}_Y^{(n)}); n=1,....N \},
\end{equation*}
where $(\tilde{\theta}_X^{(n)}, \tilde{\theta}_Y^{(n)})$ are the orientation angles of two primary particles for the $n$th observation. 

 We first looked at the angular correlation coefficients of $\tilde{\theta}_X^{(n)}$ and $\tilde{\theta}_Y^{(n)}$ for each aggregation group. Let $N_{k1,k2}$ denote the collection of observation indices $n$'s that correspond to aggregations of shape categories $k1$ and $k2$. Following \citet{fisher1983correlation}, the angular correlation coefficient $\rho_{k1, k2}$ is defined as,
\begin{equation*}
\rho_{k1, k2} = \frac{\sum_{i,j \in N_{k1, k2}} \sin(\tilde{\theta}_X^{(i)} - \tilde{\theta}_X^{(j)}) \sin(\tilde{\theta}_Y^{(i)} - \tilde{\theta}_Y^{(j)}) }{\sqrt{\sum_{i,j \in N_{k1, k2}} \sin^2(\tilde{\theta}_X^{(i)} - \tilde{\theta}_X^{(j)})}  \sqrt{\sum_{i,j \in N_{k1, k2}} \sin^2(\tilde{\theta}_Y^{(i)} - \tilde{\theta}_Y^{(j)})} }.
\end{equation*}
The corresponding coefficient of determination, $\rho_{k1, k2}^2$, is 0.1859 for Rod-Rod, 0.0273 for Rod-Ellipse, 0.0937 for Rod-NearSphere, 0.1195 for Ellipse-Ellipse, 0.0008 for Ellipse-NearSphere, 0.2252 for NearSphere-NearSphere. When $k1=k2$, the coefficients were computed in between $\min\{ \tilde{\theta}_X^{(n)}, \tilde{\theta}_Y^{(n)} \}$ and $\max\{ \tilde{\theta}_X^{(n)}, \tilde{\theta}_Y^{(n)} \}$. The small values of the coefficients imply that the two angles are weakly correlated.

\begin{figure}[h]
	\centering
	\includegraphics[width=\textwidth]{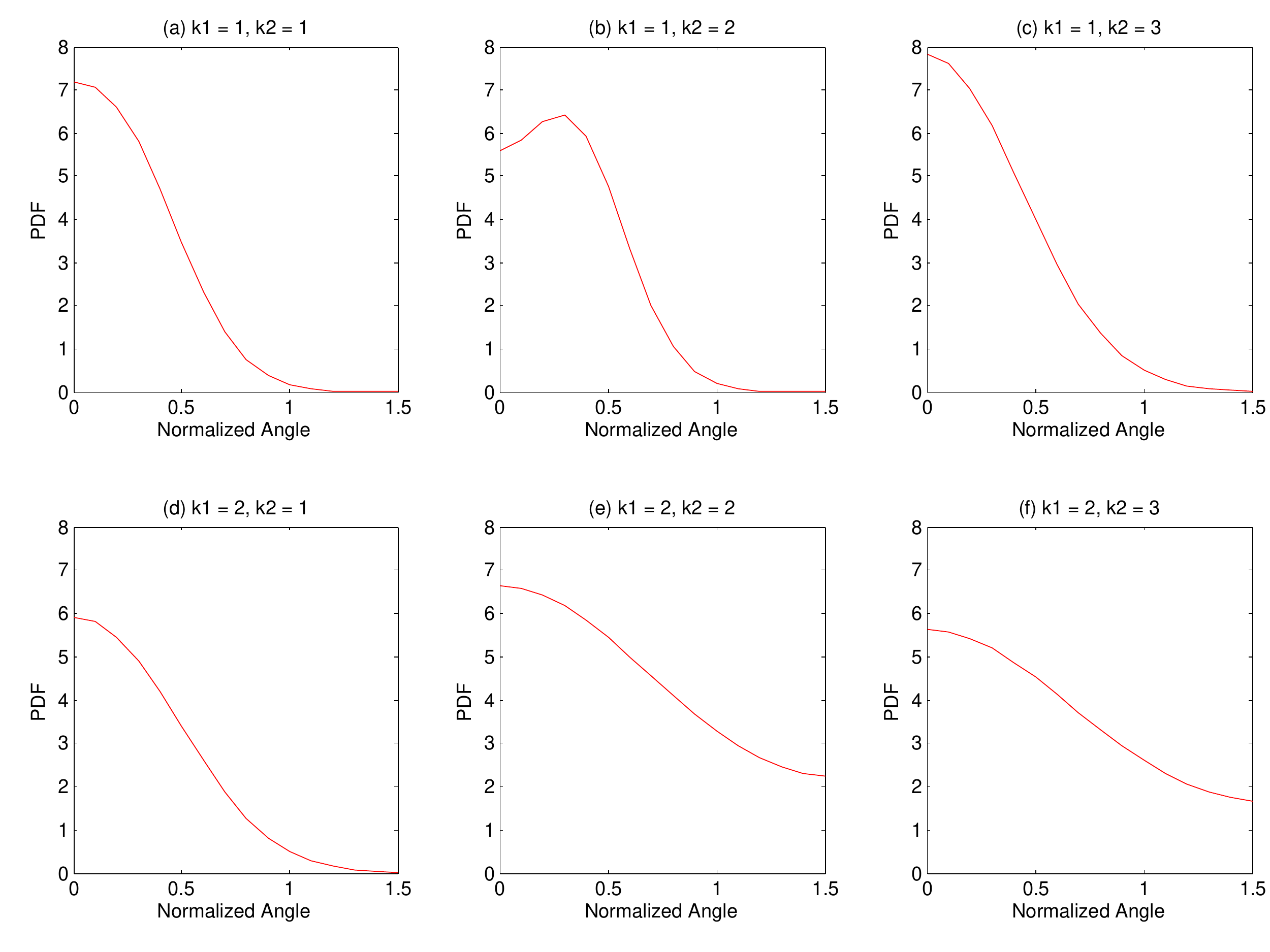}
	\caption{Estimated probability density function of the orientation of shape category $k1$ when it aggregates with shape category $k2$}
	\label{fig:pdf}
\end{figure}

\begin{figure}[h]
	\centering
	\includegraphics[width=\textwidth]{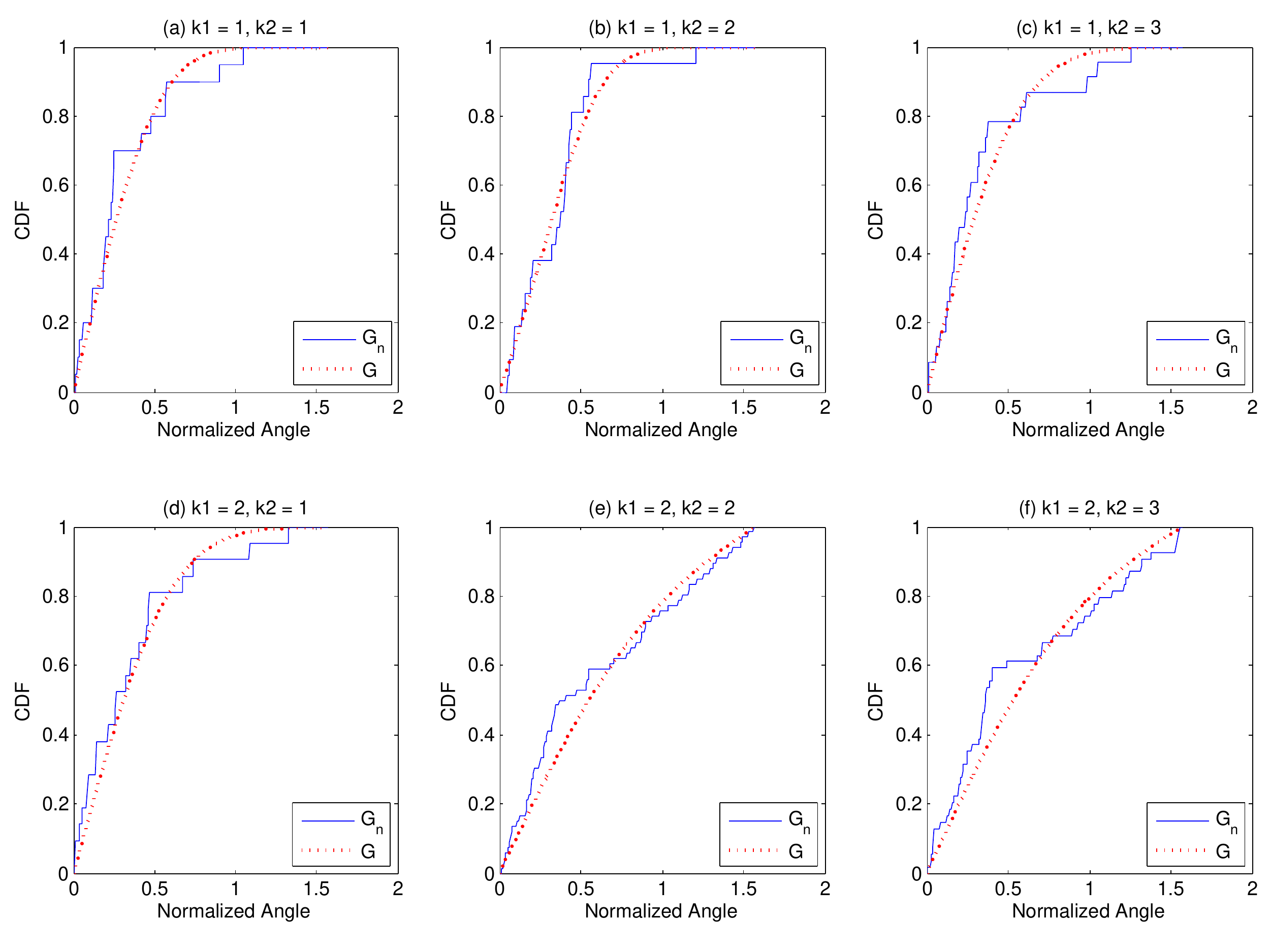}
	\caption{Goodness-of-Fit Test; $G$ denote the estimated CDF, and $G_n$ denotes the empirical CDF.}
	\label{fig:gof}
\end{figure}

Given the weak correlation of $\tilde{\theta}_X^{(n)}$ and $\tilde{\theta}_Y^{(n)}$ and a limited number of observations per group, we approximately model the joint distribution of the two angles with a product of the marginal distributions of the two angles. Let $p_{k1|k2}(\tilde{\theta})$ denote the marginal density function of $\tilde{\theta}$ of shape category $k1$ when it aggregates with shape category $k2$, which is assume to be
\begin{equation*}
p_{k1|k2}(\tilde{\theta}) = g(\tilde{\theta}; \gamma_{k1,k2}, \kappa_{k1,k2}).
\end{equation*}
The maximum likelihood estimation procedure described in Section \ref{sec:mle} was applied for $k1=1,2$ and $k2=1,2,3$. We have not analyzed $k1=3$ cases (Near-Sphere cases) because the cases are subject to relatively large estimation errors as we showed from the simulation study in Section 3.3. Let $\hat{\gamma}_{k1,k2}$ and $\hat{\kappa}_{k1,k2}$ denote the estimated $\gamma_{k1,k2}$ and $\kappa_{k1,k2}$. Figure \ref{fig:pdf} shows the $p_{k1|k2}(\tilde{\theta})$ with $\hat{\gamma}_{k1,k2}$ and $\hat{\kappa}_{k1,k2}$. 
The method described in Section \ref{sec:fit} was applied for the goodness-of-fit testing of the estimated density functions. For all cases, the estimated CDFs were very comparable to the corresponding empirical CDFs, and the goodness-of-fit test also showed no significant difference between them with 95\% significance level. Figure \ref{fig:gof} shows the cumulative density functions ($G$) that corresponds to the estimated PDFs, with comparisons to the empirical CDFs ($G_n$).  

We also tested a hypothesis related to whether there is a preferential orientation of shape category $k1$ when it aggregates with shape category $k2$. We applied the method proposed in Section \ref{sec:uniftest} to test
\begin{equation*}
\begin{split}
\mbox{H}_0: & \kappa_{k1, k2} \le 0.5 \\
\mbox{H}_1: & \kappa_{k1, k2} > 0.5.
\end{split}
\end{equation*}
With 95\% significance level, the null hypothesis was rejected for $(k1,k2) = (1,1)$, $(1,2)$, $(1,3)$ $(2,1)$, $(2,2)$ and $(2,3)$. The results indicate strong evidences that rod-like and ellipse-like nanoparticles have preferential orientations when they aggregate with rod-like, ellipse-like or near-sphere like nanoparticles. 

We performed a steered molecular dynamics (SMD) simulation of a rod-to-rod particle aggregation \citep{welch2015nature}, which allowed us to compute the energy barriers against aggregation for different orientations of rods. According to the simulation, when the major axes of two aggregating rods were not oriented toward the aggregation center, the compression of solvent monolayers at rod surfaces significantly increased when the rods became close to each other. The increase of the solvation force placed a large energy barrier against the aggregation of the two rods. The energy barrier was minimized when both of the rods' major axes were oriented toward the aggregation center. This implies the preferential orientation of a rod particle in its aggregate is zero. Note that the direction of the major axis is zero. To test how our experimental observations are consistent with the simulation result, we formulated a hypothesis testing problem, which basically examines whether the mean orientation $\gamma_{1,1}$ for a Rod-Rod aggregation is zero, 
\begin{equation*}
\begin{split}
\mbox{H}_0: & \gamma_{1, 1} = 0 \\
\mbox{H}_1: & \gamma_{1, 1} \neq 0.
\end{split}
\end{equation*}
We applied the method proposed in Section \ref{sec:meantest} to test the hypothesis. With 95\% significant level, the null hypothesis cannot be rejected. In other words, with high significance, the experimental observations are consistent with the output of the SMD simulation.

\section{CONCLUSION} \label{sec:concl}
We have presented a mathematical model for studying the oriented attachment of nanoparticles with dynamic microscopy data, i.e., studying the preferential orientations of two primary nanoparticles participating in the particle aggregation. We geometrically defined a particle aggregation by two primary geometries merging into a secondary geometry. Each primary geometry in dynamic microscopy data was represented by a simply connected subset in a two-dimensional Euclidean space with a certain choice of its standard coordinate system, and the secondary geometry was represented by a union of the two primary geometries having certain orientations. We proposed a shape alignment approach to define the orientations of the primary geometries within the secondary geometry, and presented a numerical algorithm for solving the approach. The approach was validated using simulation datasets that emulated aggregations of ellipses, since it is impossible to evaluate it for infinitely many different aggregation cases. Although the validation is limited to aggregation of ellipses, it may be applied for analyzing aggregations of other shapes in a two-dimensional image with additional validations. 

We also presented a statistical model to describe the probability distribution of the orientations of primary geometries in their aggregates and formulated several statistical hypothesis testing problems. The statistical model was specifically designed to describe a four-fold symmetric angular distribution since a four-fold symmetry was shown in our motivating example. The model was validated using simulation datasets.  

We applied our proposed method to our motivating example of nanoparticle aggregations. The results demonstrated that two primary particles were aligned along certain preferential orientations during their aggregation and the orientations were consistent with what we achieved from a molecular dynamics simulation. By far, the microscopy and nanoscience community has been manually cherry-picking and analyzing individual cases of nanoparticle aggregation. To the best of our knowledge, our study is the first attempt to statistically analyze multiple cases of nanoparticle aggregations from a single nanoparticle synthesis process.

\bigskip
\begin{center}
{\large\bf SUPPLEMENTARY MATERIAL}
\end{center}
\begin{description}
\item[Implementation details: ] a pdf file containing some implementation details of the proposed method, including Section A. the optimization algorithm to solve problems (4) and (7), and Section B. the first and second order derivatives of the log likelihood function in Section \ref{sec:mle}. 
\end{description}

\bibliographystyle{agsm}
\bibliography{AGGREGATE}
\end{document}